# Incorporating Epistemic Uncertainty into the Safety Assurance of Socio-Technical Systems

Chris Leong       Tim Kelly       Rob Alexander

Computer Science Department
University of York
York, United Kingdom
{cwkl500, tim.kelly, rob.alexander}@york.ac.uk

In system development, epistemic uncertainty is an ever-present possibility when reasoning about the causal factors during hazard analysis. Such uncertainty is common when complicated systems interact with one another, and it is dangerous because it impairs hazard analysis and thus increases the chance of overlooking unsafe situations. Uncertainty around causation thus needs to be managed well. Unfortunately, existing hazard analysis techniques tend to ignore unknown uncertainties, and system stakeholders rarely track known uncertainties well through the system lifecycle. In this paper, we outline an approach to managing epistemic uncertainty in existing hazard analysis techniques by focusing on known and unknown uncertainty. We have created a reference populated with a wide range of safety-critical causal relationships to recognise unknown uncertainty, and we have developed a model to systematically capture and track known uncertainty around such factors. We have also defined a process for using the reference and model to assess possible causal factors that are suspected during hazard analysis. To assess the applicability of our approach, we have analysed the widely-used MoDAF architectural model and determined that there is potential for our approach to identify additional causal factors that are not apparent from individual MoDAF views. We have also reviewed an existing safety assessment example (the ARP4761 Aircraft System analysis) and determined that our approach could indeed be incorporated into that process. We have also integrated our approach into the STPA hazard analysis technique to demonstrate its feasibility to incorporate into existing techniques. It is therefore plausible that our approach can increase safety assurance provided by hazard analysis in the face of epistemic uncertainty.

## 1. Introduction

Imagine a safety meeting among safety engineers, project managers and operators to evaluate the hazards affecting a system prior a flight trial. The operators raised a concern as to whether equipment item X could operate in a certain flight profile. Unfortunately, the information was not available. The equipment working procedures, which were provided during the design phase, did not include any operating specifications. While the project managers knew that the equipment operating specifications were missing, they did not anticipate that this absence required further attention after the design phase. The project managers thus did not follow up on this uncertainty. Separately, a junior engineer at the end of the table was concerned with possible distraction during the flight trial as the pilot needs to carry out multiple tasks during the flight, which was not considered during the safety meeting. Being inexperienced, he was unsure if such distraction could be safety-critical, so decided to remain quiet and not raise the issue.

To perform comprehensive safety analysis, we must be able to make timely and accurate predictions about potential hazards. Such prediction is based upon the collective wisdom and experiences of the people involved, as well as the best information available at the time of conducting the assessment. In a meeting like the one above, plausible-but-uncertain predictions or concerns may end up being discarded and ignored rather than captured and tracked. The aim of our work is to investigate if more can be done to track such uncertainty and provide better prediction regarding potential hazard during system development.

As part of the safety assurance for complicated socio-technical system (STS) [1], system stakeholders (which include multiple parties such as safety engineers, project managers, system managers and operators) capture safety-critical causal relationship so as to derive the causes of hazards. Hazards can be identified from causal relationships among entities, states, behaviours and events that are related to the system, to its surroundings, and to other systems in the STS. In this paper, we will refer to all such things as "objects". Examples of such hazards include components failure, unsafe human behaviour, unexpected software interaction, incorrect or insufficient safety practice and undesired change in external environment.

As with other activities which depend on abstractions of the real-world, hazard analysis will be affected by uncertainty. Uncertainty can be classified as aleatory or epistemic [2] – while aleatory uncertainty is random, epistemic uncertainty is due to a lack of knowledge. Our epistemic uncertainty can be due to issues we know we do not know (known uncertainties), or issues we do not know we do not know (unknown uncertainties) [3]. Although some epistemic uncertainty is unavoidable, we can minimise its undesired effects by improving the ways we manage both known and unknown uncertainties during hazard analysis. In Section 2, we elaborate on the problems of conducting hazard analysis under epistemic uncertainty. Section 3 presents our approach of capturing and tracking such uncertainty. In Section 4, we discuss the applicability of our approach. Finally, we describe the conclusion and future work in Section 5.

## 2. Issues with Epistemic Uncertainty in Hazard Analysis

In system development, epistemic uncertainty is an ever-present possibility when reasoning about the causal factors during hazard analysis. Such uncertainty is common when complicated systems interact with one another, and it is dangerous because it impairs hazard analysis and thus increases the chance of overlooking unsafe situations. Unfortunately, the problem due to uncertainty is compounded as existing hazard analysis techniques tend to ignore unknown uncertainties, and stakeholders involved in system development rarely track known uncertainties well through the system lifecycle.

### 2.1. Epistemic Uncertainty is Risky

Epistemic uncertainty in hazard analysis has a high risk of causing unsafe situations since it is common (i.e. high probability of occurrence throughout the lifecycle) and dangerous (i.e. severe enough to be safety-critical). We shall elaborate further on both observations.

- **Common**. The occurrence of epistemic uncertainty is high and unavoidable throughout a system's lifecycle. For example, during design phase, specifications and requirements may not be well defined. During acquisition phase, multiple project teams and stakeholders with different vested interest may lead to unexpected behaviours. During operation, a system may require to operate either with other systems or in an environment which has not been considered before. All the above are possible scenarios that can result in uncertainty throughout the system lifecycle.

- **Dangerous**. The presence of epistemic uncertainty can be severed enough to affect hazard analysis. Uncertainty can cause inaccurate assessment as decision makers are presented with incomplete information. Such inaccurate analysis for safety-critical system can overlook failure or risky scenarios that may result in death, injury or damage to property. In addition, uncertainty can also delay safety assessment when relevant information is not available at time of analysis.

## 2.2. Ignorance of Unknown Uncertainty

While stakeholders acknowledge the existence of unknown uncertainty, they tend to ignore it and focus on what they are aware of from their collective wisdom and experiences regarding uncertainty, that is, the known uncertainty. This is understandable during system development as there is a pressure to perform within limited resources. Stakeholders have to make risk assessment under a myriad of known uncertainties due to a lack of time, expertise and information necessary to make a good judgement. Given the limited resources, the assessment would tend to be focusing only on what is already known about the uncertainties.

However, not having the capacity to focus on unknown uncertainty does not mean unknown uncertainty is not safety-critical. We want to help stakeholders to recognise such unknown uncertainty by creating an abstract structure that encompasses possible safety-critical causal relationships for people to specify what they know and what they don't know. This is akin to the 'observability-in-depth' principle under system safety [4] to identify hazards. The principle advises stakeholders to scan the horizon for possible scenario that can transit a system to an increasingly hazardous state. In our work, we want to shift the boundary between knowing and not knowing about epistemic uncertainty, by surfacing previously unknown uncertainty during hazard analysis.

## 2.3. Lack of Tracking of Known Uncertainty

Even when there is uncertainty that we are aware of, there is still a possibility to ignore and not track it. Such information may be discarded because it could be deemed unimportant at the time it was acquired. However, uncertainty regarding any given system element can vary over time as the developer's knowledge about the system and its environment changes throughout the system lifecycle. Uncertainty can vary depending on the level of abstraction that the information is being presented. The more general the information, the greater the uncertainty. Uncertainty can also vary depending on the level of control over the system behaviour. There will be more certainties regarding a system being developed, compared to an external system or the environment that we have less control and knowledge about.

A system that is deemed simple and predictable during design phase may become complicated and uncertain when it starts to interact more with other systems. Also, some uncertainties need time before we can determine if they are safety critical. For example, there could be preliminary documents with uncertainties about operational concepts, requirements and design features that can only be validated in the later stage of a system development. If we do not track such uncertainties, we may end up losing information that may turn out to be safety-critical later. Currently, we have few or no ways of systematically and efficiently track plausible-but-uncertain causal relationships. We need a feasible and practical process to manage such uncertainty as a part of hazard analysis.

## 3. Our Approach to Manage Epistemic Uncertainty

In this section, we describe a reference, a model and a process that we have introduced to manage uncertainty. We have created a reference populated with a wide range of safety-critical causal relationships found in the literature to help recognise unknown uncertainty; and we have developed a model to systematically capture and track known uncertainty around such factors. We have also defined a process for using the reference and model to assess possible causal relationships during hazard analysis.

### 3.1. Reference of Causal Paths to Recognise Unknown Uncertainty

In safety analysis, it is expected that stakeholders may not be aware of all causal paths. Hence, we want to help them recognise causal paths that are safety-critical even though they may not have full knowledge about these causal paths. There is a lot of understanding of the nature of causal relationships from collective wisdom. In the spirit of good safety engineering practice, we want to harness the maximum effect of prior knowledge about credible causal paths. This motivates us to develop a guide to recognise plausible causal paths. Having a reference of causal factors and causal paths (we will define both of these terms in the next section) can help decision makers to identify potential hazards that can lead to unsafe situation.

To create a credible reference, we have conducted an extensive literature review of more than 30 different topics that are related to safety. While the reference cannot claim to be complete, it provides a sufficient coverage of diverse issues to help stakeholders recognise a wide range of safety-critical concerns. We have observed that as each field of study is specific to one domain within safety, none of them can serves as an isolated guide to discover all types of hazards. For example, Shappel's Human Factors Analysis and Classification System (HFACS) [5] provides a detailed review of issues related to human such as complacency, distraction and confusion; but does not focus on technology issues that can also cause uncertainty. His work can be complemented by O'Halloran's taxonomy of Failure Mode/Mechanisms Distribution (FMD) [6] that lists the possible safety-critical issues resulted from technical properties such as kinetic, chemical and electrical. In a different study, Endsley's taxonomy of situation awareness error [7] focuses on information and decision making, which provides another dimension of causal factors.

In our literature review, we started by identifying potential causal paths that may result in unsafe situations. These causal paths covered a wide range of topics such as system safety, human factor ergonomic, project uncertainty, taxonomy of safety-related subjects and situational awareness. From the list of causal paths and the suggested classifications within the literature, we have consolidated the causal paths into six primary causal factors: Human, Organisation, Technology, Process, Information and Environment. Table 1 provides a summary of the causal factors and the associated causal paths. Each of the causal factors can be further divided into two or three secondary causal factors (highlighted bold in Table 1).

The danger of over-reliance on a checklist should be emphasised here. The checklist can serve as a guide to provide reference and direction for stakeholders to recognise potential causal paths that affect safety. These will not be the only possible causal paths that can occur in a causal relationship. More importantly, the approach of considering and recognising plausible causal paths helps to shift these causal paths from being unknown uncertainty (e.g. not knowing the existence of a causal path) to known uncertainty (e.g. not having full knowledge about a causal path). This awareness of known uncertainty is better for the safety assessment than the initial state of not recognising that the plausible causal path exists.

| Causal Factors | Causal Paths |
|---|---|
| Human | **H1: Manpower** – expertise[8-10] staffing[5, 8, 10-14] mix[12] ownership[8] experience[8, 12] leadership[5, 15] skill[5, 10, 12, 13, 16-18] ability[12] characters[19] individualistic[20] demographic[20] cultural[20] obligation[21] survivable[12] stakeholders[10, 22-25] user[26] turnover[10] education[10] |
| | **H2: Mental state** – escalation[15] brokerage[15] free rider[15] convention[15] norm[15] selective benefit[15] morale and motivation [10, 12, 15] social[18, 27] deliberate[16] esteem[21] complacency[5] stress[5] overconfidence[5] fatigue[5] distraction[5, 7] confusion[5] health[12] comfort[12] visual limitation[5] illness[5] injury[12] disability[12] hearing limitation[5] cognitive[12] physical[12, 28] sensory[12] team dynamic[12, 13] aptitude[12] emotional[28] |
| | **H3: Action** – operation[9] network[15] broadcast[15] rumour[15] communication[5, 8, 10, 13] open[13] interrelation[13] atmosphere[13] engagement[29] coordination[5] omission[7, 16] commission[16] extraneous act[16] observation[19] interpretation[19] overcommit[21] performance slip[31] specification slip[31] lapse-forgot[31] lapse-overlook[31] rest[5] preparation[5] intentional violation[13, 18, 32] behaviour[29] lack involvement[10] influence[30] |
| Organisation | **O1: Management** – supervision[5, 9, 15] audit[15] communication[19] structure [5, 19, 23, 30, 33] levels of domain[30] role ambiguity and conflict[20] schedule[20] demand[21] feedback and refine[5] company[14] project size[10] project uniqueness[10] project density[10] |
| | **O2: Policy** – regulation and control[14, 15, 22, 30] job future and security[20, 21], culture and climate[5, 10, 17, 20, 33, 34] reward and recognition[20, 21] incompatible goals[10, 13, 32] trade-off[13] ambiguous goal[10] narrow goal[10] expectation[10] customer satisfaction[26] |
| | **O3: Resource** – training facility[9, 15, 19, 26, 32] material[8, 9, 17] supplier management[10, 15, 25] support facility[5, 10, 16, 26, 28] time phase[11, 16] time step[11, 16] project urgency[10] allocation[5] monetary[5, 10] instructional[12] unrealistic time frame[10] outsource management[10] interdependent |
| Technology | **T1: Machine** – hardware capability[9, 11, 18, 22, 25, 30, 32, 33] hardware compatibility[34] technical[23, 27, 35] equipment [5, 16, 19] interface[5, 19] link[18] node[18] display[5] construction[17] software[6, 11, 18, 22, 25, 30, 33, 34] communication[6, 26, 32] engineering[24] mobility[18] traffic[18] area coverage[18] services[26] tool[26] technique[26] abstraction[8] working range[8] tech change[8, 10] innovation[8] complexity[5] availability[13] function[13] |
| | **T2: Property** – energy[11] kinetic[8] biological[8] acoustical[8] chemical[8] electrical[8] mechanical[8] electro-magnetic[8] thermal[8] radiation[6, 8] bonding[6] buckling[6] change in property[6] corrosion[6] cracking[6] deformation[6] fatigue[6] seizure[6] impact[6] rupture[6] voiding[6] wear[6, 34] breakdown[6] contamination[6] diffusion[6] degradation[6] incorrect current[6] punch through[6] leak[34] loose[34] drift[34] synchronisation[34] |
| | **T3: Support** – system design[17, 32] tool design[20] tool usability[20] work area design[20] task design[5, 32] medium[18] |
| Process | **P1: Nature** – segregation[8] systematic[8] oversight[5, 8] procedure [5, 8, 11, 13, 16, 17, 19, 22, 32-34] practice[8, 22] overload[7, 20] control[11, 20] autonomy[20, 28] repetitiveness[20, 30] feedback[20, 28] ability to learn[20] input[11] output[11] lower level failure[18] cascade failure[18] delay[18] |
| | **P2: Phase** – design and plan[19, 35] validation[8] verification[8] manufacturing[24] operation[24] risk management[8, 10, 12, 32] review[8] maintenance[13, 32, 34, 35] housekeeping[32] inspection[35] supervision[35] work[14, 26, 27, 33] training[13, 16] execution and operation[5, 16, 26, 34] mis-operation[16] task[20, 23, 25] sense-making[26] decision making[26] thinking[26] |
| Information | **I1: Knowledge** – procedure[9] standard[9] method[9] assumption[16] policy[5, 25] rule[17, 22] guideline[11] precondition[11] type of info[19] manual and checklist[5] protocol[13, 18] roles and responsibilities[10] best practice[10] data[10] concept[10] no fault found[34] rationalities[30] evidence[30] values[30] fluctuation[30] customer requirements[26] codified information[26] |
| | **I2: Error** – application error[31] assumption error[31] syntax error[31] requirement error[31] lack of distinction[31] lack of awareness[31] insufficient knowledge[31] situational awareness error[13] incomplete specification[10] conflicting requirements[10] info processing problem[10, 26] data unavailable[7] |
| Environment | **E1: Physical** – transport network[15] ambient condition[16, 19] weather[5, 16] orientation[16] size[16] location[16] elevation[16] operating condition[12, 19] noise[5, 20] lighting[5, 20] vibration[5, 20] pollution[20] heat[5] terrestrial[18] meteorological[18] cosmological[18] |
| | **E2: Non-physical** – cultural[9, 26, 33] social[22] attitude[9] economic[10, 15, 18, 22, 33] competitiveness[26] political[10, 15, 18, 22, 25, 33] regulatory[26, 33] legal[10, 22] contract[15] propaganda[15] duration[16] delayed[16] alternative[21] strategic interest[21] government[14] complexity[10] security[18] |

*Table 1. Reference of causal factors and causal paths*

## 3.2. Model to Capture Known Uncertainty

To make use of the checklist from the previous section for hazard analysis, we introduce the multi-level causal relationship model and the HOT-PIE diagram, which we will elaborate in this section.

**Multi-level causal relationship model**. Causal relationships can be presented at many different levels of abstraction. Hence, we have adopted the Coleman's boat of causal pathways [15, 36] in our model so as to capture the causal relationships at multiple levels of abstraction (see Figure 1a). Coleman's model considers causation at the macro and micro level, which are commonly applied in the social and biological domains. For example, in biology, some scientists may work at the macro-ecosystem level (e.g. between human and animals) which can be highly abstract. Other scientists may work at the micro-organism level such as investigating organs and cells in the circulatory system.

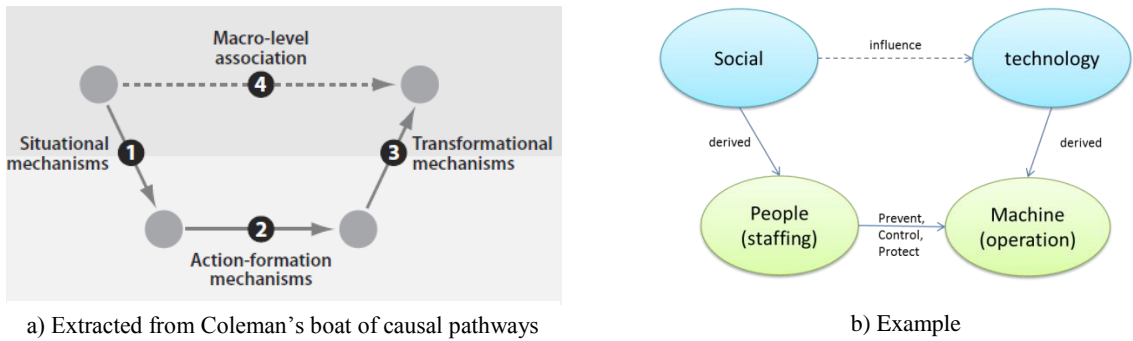

a) Extracted from Coleman's boat of causal pathways          b) Example

*Figure 1 Using Coleman's boat of causal pathway to capture causal relationships*

We can apply a similar concept when we identify safety-critical causal relationships. For example, in Figure 1b, we describe at the macro level that social factors can influence technology. We can be more precise by drilling down to the micro-level in order to show evidence of the influence of social factor on technology. One such evidence could be the lack of staffing (which is a social factor) that prevents the proper operation of the machine (which is a technological issue).

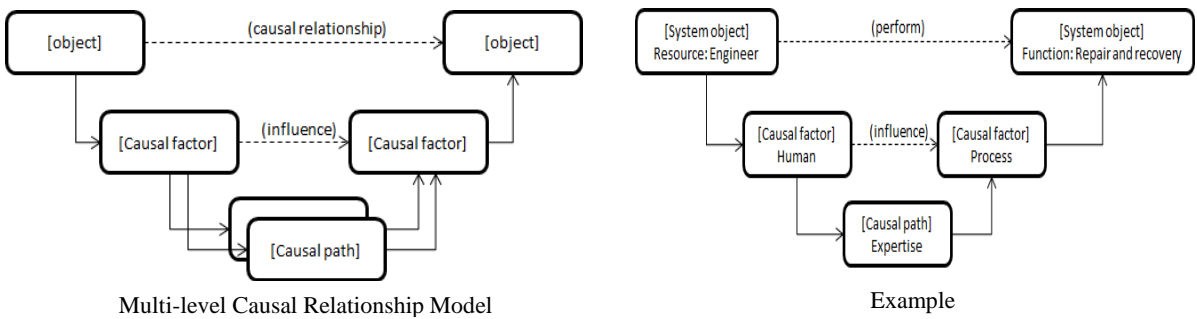

Multi-level Causal Relationship Model          Example

*Figure 2 Representing Causal Relationships in Hazard Analysis*

In our multi-level causal relationship model, we consider that a macro-level causal relationship between two objects exists when a causal factor related to one object affects a causal factor of another object. At the micro-level, these causal factors are link to each other via one or more causal paths. These causal paths are similar to the "action-formation mechanisms" under the Coleman's model.

These causal paths provide the narrative or instantiation of how objects can influence each other. Our causal model is shown on the left in Figure 2 and an example to illustrate the model is provided on the right.

In this example, the two objects are "the engineer" and "the repair and recovery function", where the former is expected to perform the latter. With reference to Table 1, we analyse a causal factor under the engineer (human) and a causal factor for the repair and recovery function (process) to identify a causal path (the level of expertise) that can potentially be safety-critical. In other words, we claim that the lack of expertise from the engineer in carrying out the process of repair and recovery may become a hazard. We can make use of Table 1 to search for other plausible causal paths between "the engineer" and "the repair and recovery function" that may be safety-critical, such as "complacency" and "performance slip".

**HOT-PIE diagram**.  Next, we introduce the HOT-PIE diagram. We have earlier defined six causal factors: Human, Organisation, Technology, Process, Information and Environment. A hexagon is used to represent these six factors that could influence or be influenced by another object. We call it a HOT-PIE diagram and is based on the first letter from each of the six causal factors (see Figure 3).

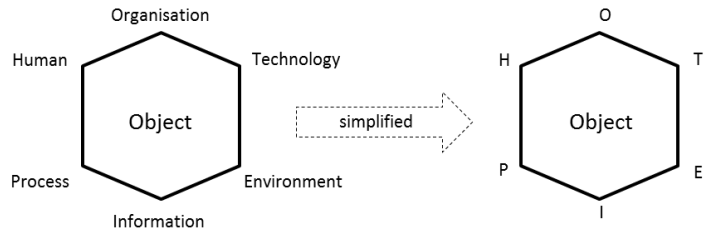

*Figure 3 HOT-PIE diagram to represent causal factors affecting an object*

An arrow connecting the vertices of two objects represents a causal path where possible safety-critical causal relationship can be derived. Back to the earlier example, the lack of expertise by the engineer in carrying out the repair and recovery function can be represented graphically using the HOT-PIE diagram. The engineer and the repair and recovery function are considered as objects, while the lack of expertise is considered as a causal path linking both objects (Figure 4).

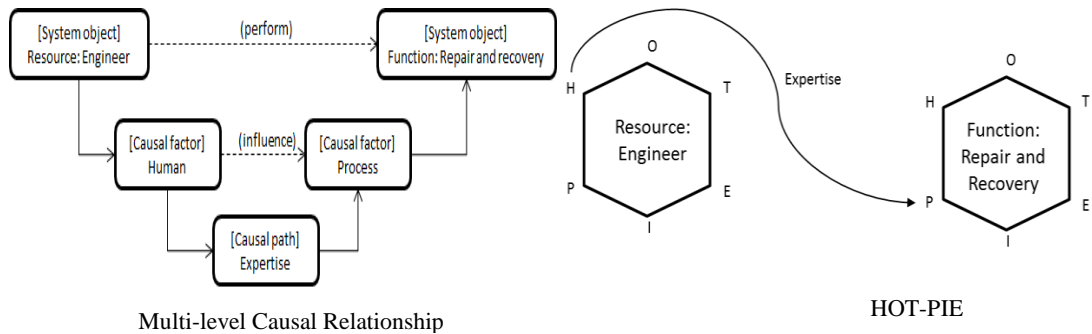

*Figure 4 Different ways of representing causal relationships*

Although the hexagonal representation in the HOT-PIE diagram may resemble the FRAM diagram [11], the foci of the two models are different. FRAM focuses on functional behaviour of a system and the hexagon in a FRAM diagram represents the six aspects of a function (time, control, input, output, resource and precondition). In contrast, our HOT-PIE diagram represents the six potential causal factors that can influence different objects and the vertices are linked by potential causal paths. While FRAM is used to analyse functions, the HOT-PIE diagram can be applied to multiple system models, such as functional, structural and behavioural.

The HOT-PIE diagram is useful when one wants to capture causal paths between system objects, such as from a design document during a hazard analysis. The diagram is simple to understand as it is based on the six causal factors. It also presents an easy way to document potential causal paths between objects. Even if we are not confident enough to conclude that a causal path is indeed safety-critical, we can still capture the potential causal path easily for future analysis when the relevant information is available.

### 3.3. Process to Augment Hazard Analysis Technique

Instead of a separate standalone method to conduct safety assessment, our approach aims to incorporate the considerations of epistemic uncertainty through augmenting existing hazard analysis techniques such as the STPA[1] [37] and FMEA[2] [38]. By introducing complimentary steps to existing analysis, we make the safety assessment more complete through recognising the influence of uncertainty on safety-critical causal paths. Our checklist of causal paths helps stakeholders to recognise potential causal relationships and the HOT-PIE model can be used to consider the various causal factors related to each object. The desired outcome is to recognise unknown uncertainties, capture them as know uncertainties and track them throughout the system lifecycle. We summarise the three steps in Figure 5.

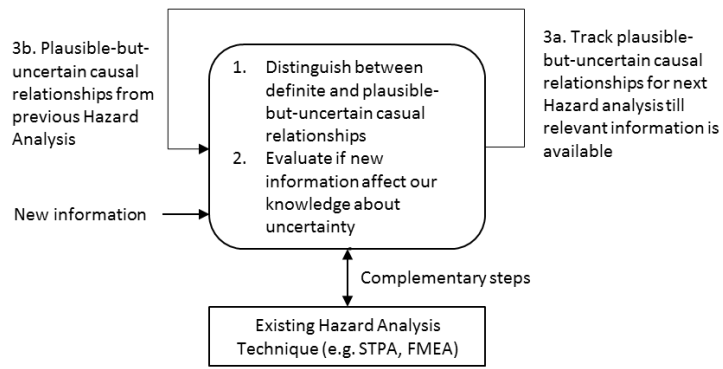

*Figure 5 Tracking of epistemic uncertainty through existing hazard analysis technique*

**Step 1: Recognise Definite and Plausible Causal Relationships**. With the help of the causal paths checklist, we aim to recognise previously unknown causal relationships affecting the STS. Some of these may turn up to be safety-critical and considered as hazards. Others may not be considered as safety-critical during the analysis due to uncertainty (e.g. not knowing if the eventual engineer doing the runway repair has the necessary training and qualification). Instead of considering all causal relationships the same, we differentiate those that are plausible-but-uncertain from those that are more definite. For causal relationships that are specific and definite, we can immediately make use of them as evidence during hazard analysis. On the other hand, for those plausible-but-uncertain causal relationships, we want to capture them for future analysis.

**Step 2: Evaluate New Information Affecting Uncertainty**. We have earlier mentioned that uncertainty can evolve and hence, it is important for us to track it. Some uncertainties will become clearer with time, such as the availability of test result, confirmation of the actual engineer for the runway repair and the availability of interface specifications of yet to be developed component. Such relevant information will enable us to make better judgement about safety risk, albeit at a later phase of the system lifecycle. To enable this, we need to rigorously track uncertainties till the relevant information is available down the lifecycle. We want to encourage stakeholders to defer judgement on potentially safety-critical causal relationships, rather than always discard them, as we may not fully appreciate their influence in a complicated STS at the point where we first considered them.

---

[1] STPA (Systems-Theoretic Process Analysis) is a hazard analysis technique based on Systems Theory.
[2] Failure Mode and Effects Analysis is an inductive reasoning technique for hazard analysis.

**Step 3: Enable Through-life Tracking**. We see potential of through-life tracking by complementing current safety case development. A safety case is a "structured argument, supported by a body of evidence that provides a compelling, comprehensible and valid case that a system is safe for a given application in a given operating environment" [39]. Safety cases have been widely adopted across many industries including defence, aerospace, automobile and railways. While a safety case provides a systematic structure to capture arguments that may concern epistemic uncertainty, it is often conducted at the tail end of a system development, during deployment or operation. At this point, the developing system cannot be readily modified in response to the new safety concerns. We have mentioned that epistemic uncertainties are common and unavoidable throughout the system lifecycle. For example, there could be design documents with uncertainties about operational concepts and requirements. These are not considered in safety cases if they are not developed right from the design phase. Hence, it is sensible to extend safety case to early design and track it through the lifecycle. Our approach of tracking epistemic uncertainty will be useful to complement such dynamic safety case development by incrementally tracking the impact of uncertainty. In the next section, we will show with an example how through-life capturing and tracking of causal paths can lead to better assurance later in the system lifecycle.

## 4. Applicability of Our Approach

We present three examples to assess the applicability of our approach. First, we analyse different types of system model to identify if there are causal factors that are not apparent in each model. We have chosen the U.K. Ministry of Defence Architectural Framework (MoDAF) for our analysis as it is widely used for system development. Secondly, we have also reviewed an existing safety assessment example to determine if our recommended process could be incorporated in the analysis. We have chosen the Aircraft System analysis in ARP-4761 [40] as the guideline is considered an acceptable means of establishing assurance process for aircraft system. In our last example, we use the STPA process to illustrate the feasibility of integrating our approach into existing hazard analysis technique.

*Figure 6 Summary of MoDAF viewpoints*

## 4.1. Ministry of Defence Architectural Framework (MoDAF)

MoDAF is a commonly used architectural framework that comprises multiple models or views to describe a military STS. The type and category of viewpoints are extracted from the MoDAF handbook[41] and presented in Figure 6. As hazards are mostly identified at the operational and system perspectives, we have correspondingly narrowed our analysis to operational and system viewpoints. In addition, we have also focused on the structural and behavioural categories as they are the more common models that are used for hazard analysis. Hence, we have narrowed our analysis on five operational views (OV-2, 4, 5, 6 and 7) and five system views (SV-1, 2, 4, 10 and 11) as highlighted in Figure 6. For each of the ten views, we analyse the type of data objects that can be represented and compared them with the six causal factors we have defined: Human, Organisation, Technology, Process, Information and Environment. We want to analyse the extent that causal factors are being considered in each view.

As an illustration, consider SV-1 (System View 1 – Resource Interaction Specification). SV-1 specifies the composition and interaction of resources, which can be physical artefacts, software or human resources. The key data objects related to SV-1 are extracted from the MoDAF handbook and shown in Figure 7. Using this information, we analysis if each causal factor and its associated causal paths are being considered by the data objects. The observation is summarised in Table 2. We have observed that organisation, technology and information causal factors are mostly represented in SV-1, while environmental factors are not. In addition, human factors are only partially represented as although SV-1 can show manpower deployment, it does not represent human mental states. Process factors are also partially represented as SV-1, being a structural model, needs a corresponding process model (e.g. SV-4) to better represent causal paths related to processes.

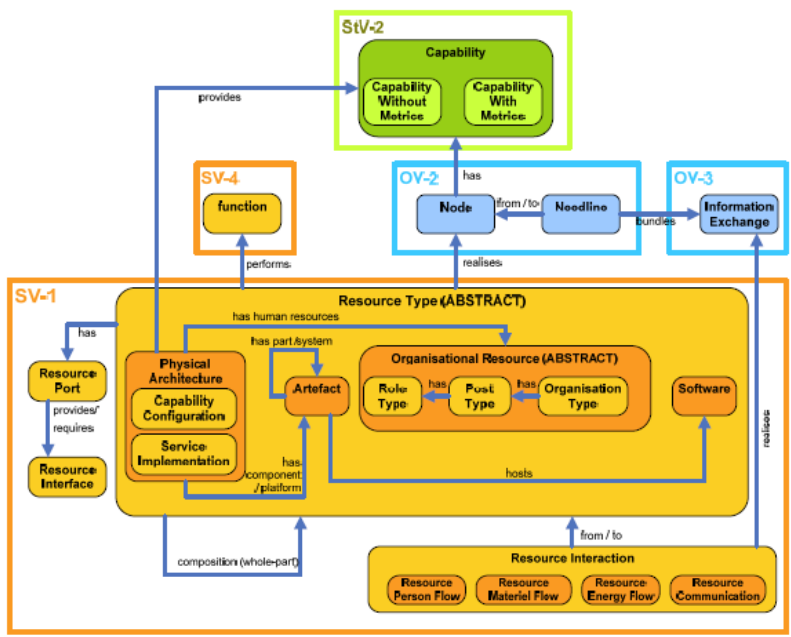

*Figure 7 SV-1 Resource Interaction Specification*

| System views | Category | Causal factors | | | | | |
|---|---|---|---|---|---|---|---|
| | | Human | Organisation | Technology | Process | Information | Environment |
| SV-1: resource interaction specs | Structural | Partially represented | Represented | Represented | Partially represented (analysed together with SV-4) | Represented | Not represented |

*Table 2 Extent of representing causal factors in MoDAF SV-1*

We have carried out similar analysis for the remaining 9 operational and system views (see Table 3). From the table, we can observe the following:
1. Different causal factors and causal paths are better represented by different views.
2. Environment related causal factors are not explicitly represented in most of the views.
3. Even if a type of causal factor is described in a view, not all the causal paths related to a causal factor are considered. For example, human related causal paths are so wide that not one view can fully represent all of them.

| System views | Category | Causal factors | | | | | |
|---|---|---|---|---|---|---|---|
| | | Human | Organisation | Technology | Process | Information | Environment |
| SV-1: resource interaction specs | Structural | Partially represented | Represented | Represented | Partially represented (analysed together with SV-4) | Represented | Not represented |
| SV-2: system communication description | Structural | Not represented | Not represented | Partially represented (only those related to communication) | Not represented | Represented | Not represented |
| SV-11: physical schema | Structural | Not represented | Not represented | Partially represented (only those related to communication) | Not represented | Represented | Not represented |
| SV-4: functional description | Behavioural | Partially represented | Not represented | Represented | Partially represented (analysed together with SV-1) | Represented | Not represented |
| SV-10: resource constraints | Behavioural | Partially represented | Represented | Represented | Represented | Represented | Partially represented |

| Operational views | Category | Causal factors | | | | | |
|---|---|---|---|---|---|---|---|
| | | Human | Organisation | Technology | Process | Information | Environment |
| OV-2: Operational node description | Structural | Partially represented | Represented | Represented | Partially represented (analysed together with OV-5) | Represented | Not represented |
| OV-4: organisational relationship chart | Structural | Partially represented | Represented | Not represented | Partially represented | Not represented | Not represented |
| OV-7: information model | Structural | Not represented | Not represented | Partially represented (only those related to info exchange) | Not represented | Represented | Not represented |
| OV-5: operational activity model | Behavioural | Partially represented | Not represented | Represented | Partially represented (analysed together with OV-2) | Represented | Not represented |
| OV-6: resource constraints | Behavioural | Partially represented | Represented | Represented | Represented | Represented | Partially represented |

*Table 3 Extent of representing causal factors in MoDAF operational and system views*

From our analysis of the MoDAF views, we conclude that none of the individual views can represent all the causal factors. This means that each view does not allow the user to fully comprehend the danger posed by every causal factor during hazard analysis. For example, using a system structural view (SV-1) may not help to identify the hazards associated with human behaviour. Similarly, a concept of operation under the operational activity model (OV-5) will not be able to surface organisation issues such as training or manpower constraints.

Our approach of managing uncertainty in causal relationships can complement MoDAF by highlighting causal factors and causal paths that are potentially safety-critical. The HOT-PIE model can help to sieve out unknown uncertainties which may not be obvious in each of the MoDAF views. These causal paths can be used either to compare with other MoDAF views or for future hazard analysis.

## 4.2. ARP-4761 – Aircraft System Analysis

SAE ARP-4761 is an industrial standard for conducting safety assessment process to certify civil aircraft. It includes a worked example of a typical safety assessment process for a fictitious aircraft design. We have studied this example to assess whether our approach could be integrated into it.

We focus on the aircraft functional hazard analysis (FHA) as it is one of the safety processes where we identify hazards. There are many aircraft functions to be investigated and we have narrowed our analysis to the 'Decelerate aircraft on the ground' function. That is the ability of the aircraft to decelerate and stop safely when it touches down on the runway. In the example, the following possible failure conditions and assumptions related to the aircraft were determined (see Table 4).

| Functional Failure Conditions: | Environmental and Emergency Configurations and Conditions | Applicable Phases: | Interfacing Functions: |
|---|---|---|---|
| a. Loss of all deceleration capability | a. Runway conditions (wet, icy, etc.) | a. Taxi | a. Air/Ground Determinations |
| b. Reduced deceleration capability | b. Runway length | b. Takeoff to rotation | b. Crew Alerting (Crew warnings, alerts, messages) |
| c. Inadvertent deceleration | c. Tail/Cross wind | c. Landing Roll | |
| d. Loss of all auto stopping features | d. Engine out | d. Rejected takeoff (RTO) | |
| e. Asymmetrical Deceleration | e. Hydraulic System Loss | | |
| | f. Electrical system loss | | |

*Table 4 Aircraft system failure conditions and assumptions*

**Applying our Approach to the Aircraft System Analysis**. During the FHA, the aircraft function tree was used in the analysis (see Figure 8). This is analogous to the SV-4 view (system functional description) in a MoDAF model. We can refer to Table 3 and find out where are the possible causal factors that may not be obvious in such a functional representation. From the row in Table 3 that describes a SV-4 view, we can generally expect that human, organisation and environment factors will not be well represented in an aircraft function tree.

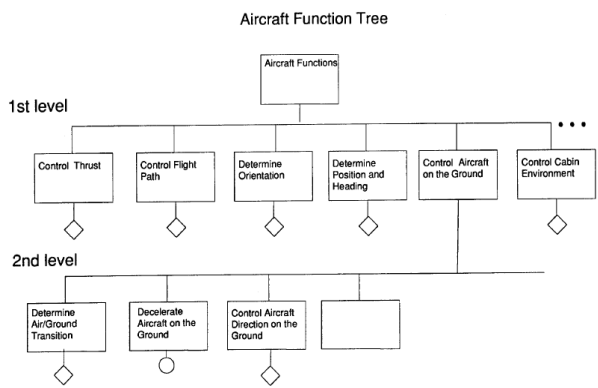

*Figure 8 Example of aircraft functions*

We have identified five top-level objects for the aircraft system analysis: aircrew, ground crew, aircraft technical systems, runway and the environment. Next, we apply the HOT-PIE causal factors on these objects to search for potential causal paths. Using the HOT-PIE diagram and referencing the checklist of causal paths, we have identified three interesting causal paths that are important to the aircraft system analysis but are not obvious in the ARP-4761 example (see Figure 9). Details of the causal paths are given in Table 5.

| CP1 | Causal Factors: Human (aircrew), Process (a/c tech system) |
|---|---|
| | Causal Path: Distraction |
| | Scenario: Pilot may be distracted due to bad practices during the deceleration process. |
| CP2 | Causal Factors: Environment (environment), Technology (a/c tech system) |
| | Causal Path: Adverse weather - hydroplaning |
| | Scenario: Wet runway may cause hydroplaning in the autobrake system, which may result in the autobrake sensor not detecting aircraft touchdown condition. |
| CP3 | Causal Factors: Organisation (ground crew), Process (runway) |
| | Causal Path: Inadequate training for runway emergency management |
| | Scenario: Unsure if adequate training has been provided to the ground crews (e.g. air traffic controllers, ground logistics team, ground runway emergency team) in preparation for adverse weather operation and emergency. |

*Table 5 Potential Causal paths between aircraft system objects*

CP1 (distraction) and CP2 (hydroplaning) are safety-critical, and they can directly affect the aircraft design. They should be fed back into the FHA and treated as potential hazards for follow-up safety assessment. CP3 concerns the qualification of ground crews in handling emergency during operation. During the high level aircraft FHA, the team involves in the analysis may not have the relevant information regarding the ground crews and it may want to focus on issues directly related to the aircraft design. In our approach, we propose to consider CP3 as a plausible-but-uncertain causal relationship. This shall be tracked through the lifecycle as long as we are uncertain if it is safety-critical. There could be many ways that CP3 can evolve as we gain more knowledge about the quality of training for the ground crews. For example, there may eventually be confirmation that the ground crews subscribe to the standardised ICAO Global Runway Safety Programme. If so, CP3 will not be of safety concern. Alternatively, it may be revealed during the system validation phase that the procedure used by the ground crew for emergency handling is different from that being used by the pilot. This can become a potential hazard affecting the aircraft landing.

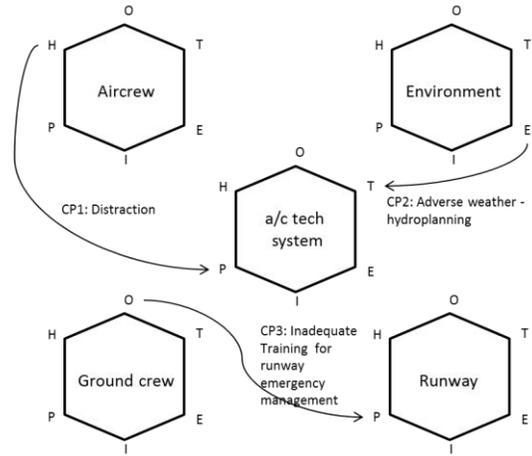

*Figure 9 HOT-PIE diagram for aircraft system analysis*

## 4.3. Augmenting existing STPA Hazard Analysis

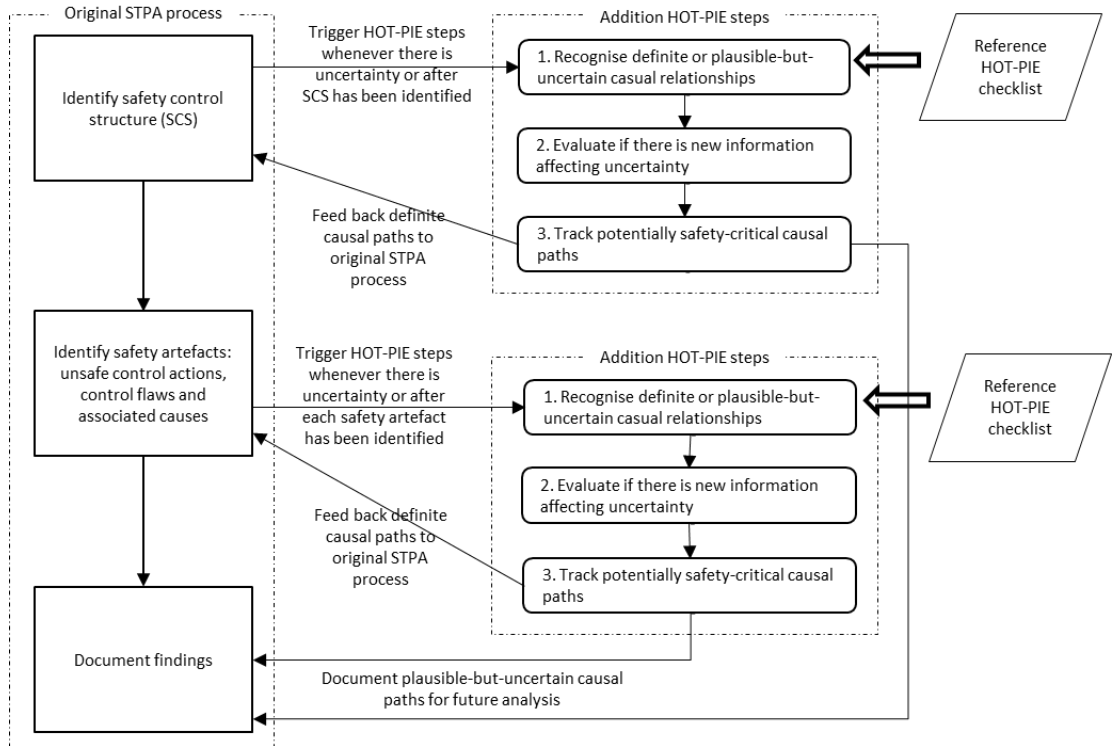

*Figure 10 Augmenting STPA with HOT-PIE approach*

As mentioned in section 3.3, we are not proposing a separate standalone method to conduct safety assessment but rather introduce additional steps within the existing hazard analysis techniques. In this section, we illustrate the use of our approach to augment the STPA process.

Figure 10 shows how our approach can integrate with STPA (details of the original STPA process is described in the STPA primer [37]). On the left are the three key steps in STPA. Our HOT-PIE approach can be introduced in the first two STPA steps to identify uncertainties among the safety control structure, unsafe control actions and control flaws. STPA requires that a system design be available; using our approach, stakeholders can reference this system design as a basis to identify causal paths concerning the six primarily HOT-PIE causal factors. Causal paths discovered from using the HOT-PIE checklist that are definite to be safety-critical will be fed back to the STPA process. Causal paths that the stakeholders may not know enough due to uncertainty would be documented as findings under the STPA process.

## 5. Conclusion and Future Work

In this paper, we have outlined an approach to managing epistemic uncertainty in existing hazard analysis by creating a technique for recognising unknown uncertainty and developing a model to systematically capture known uncertainty. To assess the applicability of our approach, we have analysed the widely-used MoDAF architectural framework and determined that there is potential for our approach to identify additional causal factors that are not apparent from individual MoDAF views. We have also reviewed a portion of the ARP4761 Aircraft System FHA example and determined that our approach could indeed be incorporated into a process like that. To further demonstrate its practicality, we have integrated our approach into STPA hazard analysis technique. It is reasonable to conclude that our approach can increase the safety assurance during hazard analysis in the face of epistemic uncertainty.

Our model provides a systematic approach to consider the effect of multiple causal paths affecting the safety of complicated system. By creating the awareness of what we know and what we don't know, it encourages stakeholders to be disciplined and explicit about the level of information and uncertainty encountered during hazard analysis. Our approach highlights (1) causal paths that are considered openly during the hazard analysis, (2) causal paths that are considered intrinsically, which may not be visible in existing hazard analysis techniques, and (3) causal paths that are unknown initially.

By advocating the capture of plausible-but-uncertain causal relationships, we have created the flexibility to defer part of the hazard analysis. This is possible by tracking the known uncertainties for future assessment till the relevant information is available. It may well be that the HOT-PIE approach reveals that certain form of causal interactions cannot yet be revealed, given the extent of the knowledge available at this point of the lifecycle. Our approach allows us to appreciate this incompleteness when augmenting existing hazard analysis techniques. We believe that this provides better assurance than an approach which claims undue confidence that we can know for sure the severity and criticality of a hazard, especially in early life cycle.

One extant concern is how well the HOT-PIE model can scale to larger systems with many objects and causal paths. One possible research area concerning large-scale application of HOT-PIE approach is to automate the process of capturing the causal factors (e.g. input into a spreadsheet via a user form). Another follow-on task is to derive criteria to assess the significance of introducing our approach to existing hazard analysis. Potential criteria include the number of safety-critical causal paths identified and the number of additional steps needed in the analysis when using our approach.

We have briefly discussed our efforts to integrate with safety case development. One option is to track causal paths as safety artefacts (e.g. evidence and arguments), which are familiar terms in safety case development. Using the work by Hawkins [42] on confidence argument, it may be possible to incorporate epistemic uncertainty in confidence arguments. This will help to support through-life tracking of safety assurance case during system development.